# Characterization of a Biosensor Based on Graphene Field Effect Transistors for Body Fluid Analytes – Channel Resistance

Bravo R[1], Silva R[2], Barret E[2], Brunnings J[2], Segarra A[3]

1- Faculty, Biomedical Engineering Department – Politechnic University of Puerto Rico.
2- Staff, Vitasenti Corporation
3- Senior Student, Biomedical Engineering Department – Politechnic University of Puerto Rico.

**Abstract** – Field-Effect Transistors with graphene channels or GFETs are an interesting alternative for the detection of analytes in biological fluids since the electrical behavior of the channel changes when exposed to a sample (among other detection strategies). In this work a preliminary characterization is made in terms of the resistance of the channel for a commercial device that has GFETs of 1 and 3 channels for cases of dry and wet gate at atmospheric pressure. The channel resistance was obtained by sweeping the drain-source voltage from -1 to +1V and measuring the drain current in a test station developed for this purpose, for gate cases with and without a PBS 0.001X reference solution. The ohmic response of the channel is linear current with respect to voltage, being greater resistance in the case of wet gate. An increase in resistance with respect to voltage was observed that it is important to review. It was possible to make the ohmic characterization of the channel and a series of recommendations are suggested to continue this research.

## 1 INTRODUCTION

Graphene channel field effect transistors (GFETs) has been used in recent years as a sensor solution in several interdisciplinary domains of biomedical engineering (Rondón et al., 2024) as an alternative, fast, compact and cost-effective method for the detection of substances and molecules (analytes) in body fluids, ranging from simple compounds to viruses (Banerjee S, 2020).

Conventional detection by GFETS (Banerjee S, 2020) is based on the continuous current (DC) analysis of the electrical response between drain and source when the graphene channel has been filled with a test substance, describing the ohmic response of the channel and the location of the Dirac voltage (gate driving threshold voltage). It has been possible to associate the presence of certain analytes in the channel with electrical responses in the aforementioned values, however, the resolving power is still insufficient for some critical analytes (viruses, for example) compared to traditional clinical laboratory methods (Chircov C, 2020), which are slow, expensive and complex (Yue W, 2017).

In the transistor gate the analytes are placed in solution with a base that is usually physiological saline, but depending on the size of the solution droplet, the higher the volume, the greater the gate capacitances and the measurement loses effectiveness, but small droplet sizes, they may not have enough analyte in solution. To avoid this problem, it is proposed to modify the concentration of sodium chloride in the base solution and observe the variations of the DC response between drain and surce, for equal volumes of the sample solution droplets.

It is proposed in this research paper to compare the response between drain and source in terms of channel resistance measured in a commercial GFET biosensor for different concentrations of base solution at equal volumes and device configuration. This characterization is the basis of the definition of the "normal" sample condition which wil be the reference for experimental test samples.

## 2      METHODOLOGY

In general, the methodology consists of the selection and provision of the device and consumables, then perform electrical tests of ohmic characterization of the channel in the presence or not of reference solution, the tabulation and analysis of results and finally the conclusions and recommendations.

*Devices and Solutions*

For the experimental phase, a commercial GFET device model mGFET-4P (Graphenea Inc. San Sebastian, Spain) provided by Vitasenti was used, and whose layout can be seen in Figure 1. This device is already encapsulated in a 20-pin JEDEC-C package and has integrated 7 single-channel GFETs and 7 triple-channel GFETs (figure 2). For this first phase of the research reported in this work, a single-channel transistor and a triple-channel transistor were randomly selected.

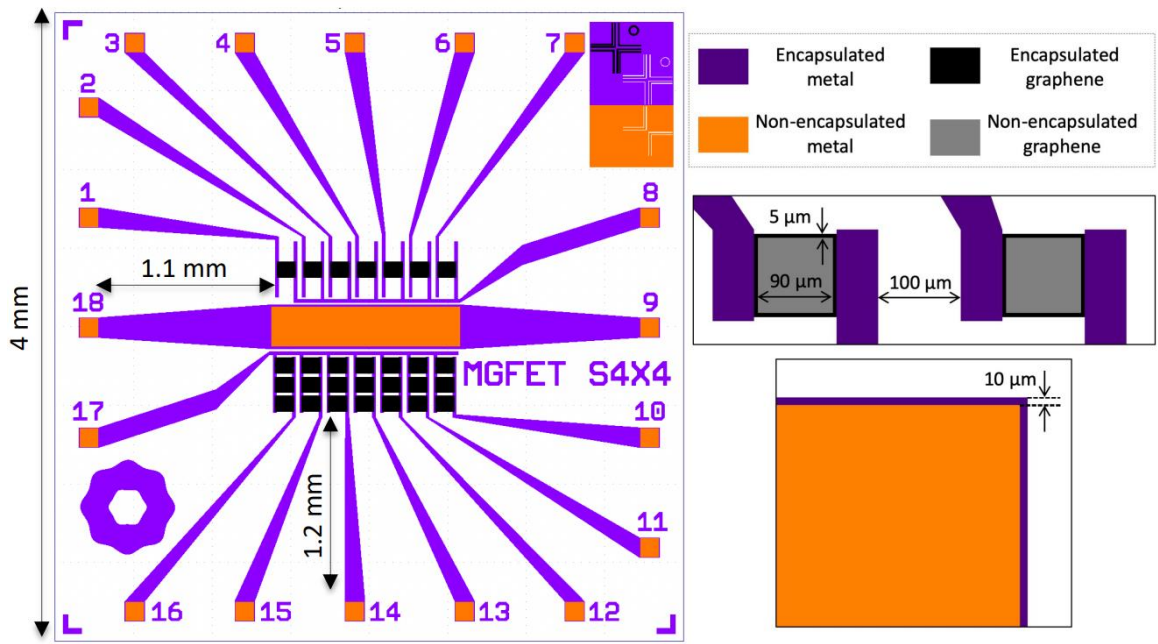

Figure 1. Layout of GFET used: Graphenea mGFET-4P (Courtesy of Graphenea Inc.).

The test solution selected was PBS 0.001X (Phosphate Buffered Solution) as it is recommended by the manufacturer for wet characterization purposes.

*Electrical Tests:*

GFET no. 2 (single channel) and no. 12 (triple channel) were randomly selected. See Figure 2-b, "U" numbering starting in the left GFETs column, upper element, counterclockwise. For the electrical tests, a small workstation based on an smart breadboard (Digilent Electronics Explorer figure 3- above) and a chassis with terminals were developed to allow future connection to automated measuring instruments (figure 3-below).

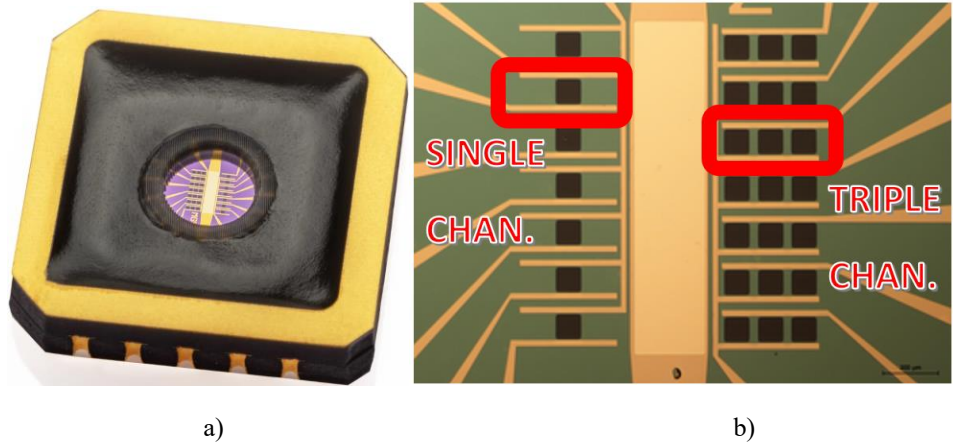

Figura 2. Physical presentation of the used device: a) commercial casing, b) Single channel GFETs and triple channel GFETs. (Courtesy of Graphenea Inc.)

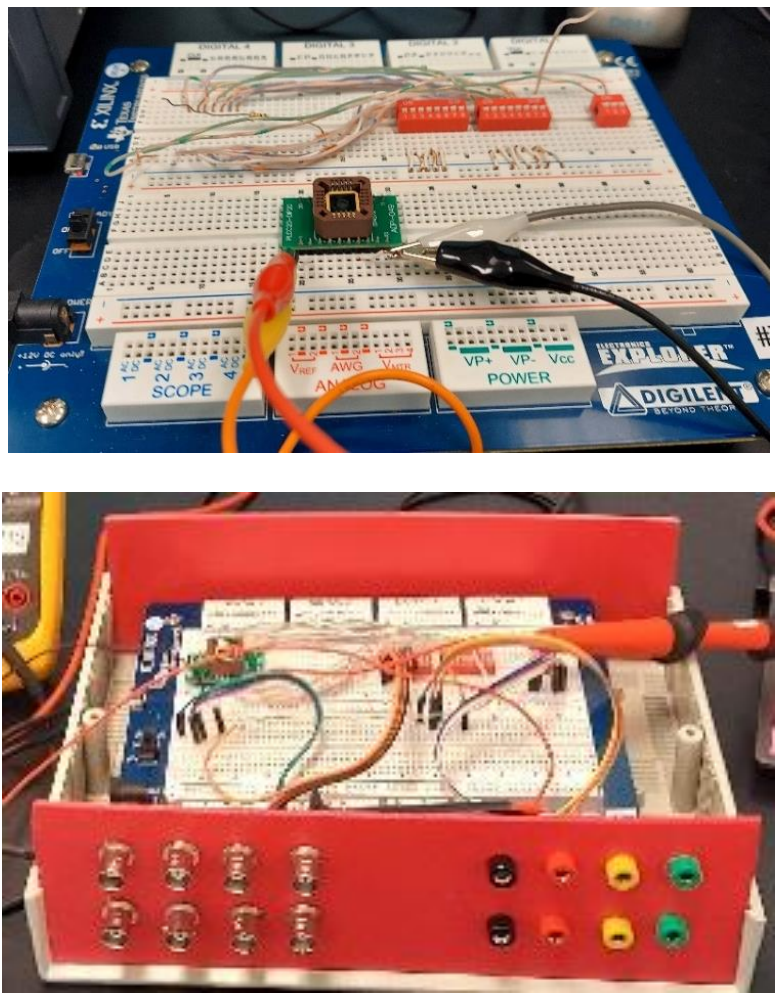

Figure 3. Workstation for electrical testing. Top – smart breadboard Digilent Electronics Explorer. Down – chassis with frontpanel of connections for instruments. (Own elaboration)

The workstation allows to assemble in a practical way the test circuit given by the following schematic (figure 4):

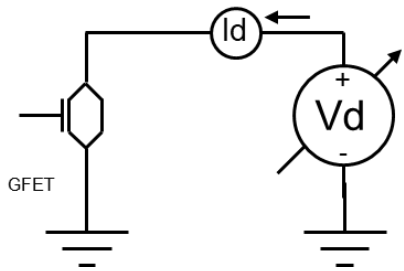

Figura 4. Test schematic circuit. (Own elaboration)

A drain source test voltage (Vd) is applied, and a test drain current Id. is measured. The voltage sweep will be done in one direction using Vd from -1 to +1V in steps of .1V (20 samples). The tests carried out below were done at the request of the research partner ally.

The physical implementation of the setup for testing can be seen in Figure 5 below:

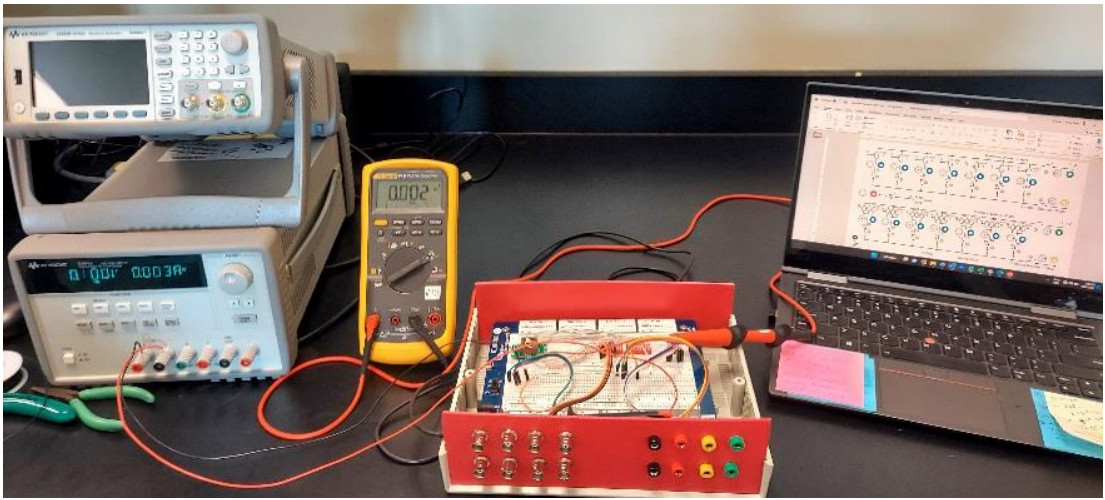

Figure 5. Setup for electrical tests. (Own elaboration).

The first set of tests consisted of characterizing the channel with the dry gate, at atmospheric pressure for the single channel transistor and the triple channel transistor and comparing it with the manufacturer's data.

The second set of tests consisted of measuring the channel resistance for the single channel transistor with the wet gate, carrying a drop of PBS 0.001X solution and comparing the results.

*Handling of GFETs*

Transistors are kept in vacuum storage while they are not being used for testing. For that, a conventional vacuum food sealer was used, as an economical alternative for that purpose.

At the end of each day of testing, the transistor gate was blown with a nitrogen jet (to clean and/or dry some material residue in the gate), then a drop of acetone was left for 8 hours. Again a nitrogen blow was applied and stored under vacuum as described in the previous paragraph.

## 3    RESULTS AND ANALYSIS

Once the measurements indicated in the methodology were made, the results were tabulated (Table 1 and Table 2) and plotted for comparison and analysis purposes (Figures 6 and 7).

In table 1, for the columns of the middle and right, the experimental ones, it is observed that for 0V a current appears, this is possible that it is because, although the voltage indicators of the source marked 0.000V, there was no precision on orders of hundreds of microvolts, that is, fractions of millivolts that could justify such a measurement.

In Table 2, in the case of wet gate, it was obtained that both voltage and current measurement instruments marked 0 in all their significant figures of the scale, so an indeterminate value was obtained that is discarded from the statistics.

It is observed in all the measurements of the tables a very slightly nonlinear behavior, increasing with voltage, that is, the resistance increases with voltage, therefore the conductance decreases and the current presents a decreasing behavior, but minimal (all the goodness of adjustment for the data have $R^2 = 1$, except for the wet channel that is 0.99, still very high).

This is a behavior observed by other researchers (Rodriguez, 2013) and that is a phenomenon that is still being considered in GFET models and that does not have much effect on circuit designs that use GFETs as amplifyer or switching devices (Rodriguez, 2013).
However, in this work, it is a behavior to which adequate attention must be paid since precisely the ohmic characteristic of the channel is the basis of the detection of analytes.

For the wet gate case, the deviation is greater (compare the variances), which adds importance to the finding because precisely the search for analytes presupposes a gate with a liquid solution that functions as a vehicle for the samples (Mackin C, 2018).

When comparing the channel resistances between single-channel GFETs vs. triple-channel GFETs (Table 1 and Figure 6), a decrease in the equivalent resistance is observed for the triple-channel case, which is expected given that resistive channels are placed in parallel that in total verify a lower resistance.

However, it is noted in the chip layout (Figure 1) that each triple device has three channels of individual GFETs in parallel, so if each individual channel is assumed to have $R_{chan}$ resistance, the triple-channel GFET should have an equivalent resistance of $R_{chan\ triple} = R_{chan}/3$ theoretically.

What the data report for the simple channel (Table 1, middle column) is that $R_{chan} = 1{,}875.031 \pm 5.173\ \Omega$ and for the triple channel (Table 1, right column) $R_{chan\ triple} = 1{,}087.631 \pm 3.294\ \Omega$, making the quotient of both, $R_{chan\ triple} = R_{chan}/1.723$ and not $R_{chan}/3$ as expected.
When comparing the results of the manufacturer's data with the experimental measurements for the single channel, it is observed that the resistance value provided by the manufacturer is lower, however, this measurement was performed under vacuum which could explain such a difference.

The tests carried out here constitute a good proof of concept with limited time and resources, only one GFET of each type was taken and only one voltage run from lower to higher, so the observations made must be corroborated and complemented with a repeatability study, an aspect that is important to consider in the continuation of this work.

**Table 1. Channel Resistance Measurements**
Drain currents vs. Vds for single channel transistors and triple channel transistors, and provider data
(Source: authors)

Provider SINGLE

| V(V) | I(mA) | R(ohm) |
|---|---|---|
| -1 | -1.25 | 800 |
| -0.9 | -1.125 | 800 |
| -0.8 | -1 | 800 |
| -0.7 | -0.875 | 800 |
| -0.6 | -0.75 | 800 |
| -0.5 | -0.625 | 800 |
| -0.4 | -0.5 | 800 |
| -0.3 | -0.375 | 800 |
| -0.2 | -0.25 | 800 |
| -0.1 | -0.125 | 800 |
| 0.0 | 0 | 800 |
| 0.1 | 0.125 | 800 |
| 0.2 | 0.25 | 800 |
| 0.3 | 0.375 | 800 |
| 0.4 | 0.5 | 800 |
| 0.5 | 0.625 | 800 |
| 0.6 | 0.75 | 800 |
| 0.7 | 0.875 | 800 |
| 0.8 | 1 | 800 |
| 0.9 | 1.125 | 800 |
| 1 | 1.25 | 800 |

| R prom | 800.000 |
|---|---|
| R std | 0.000 |

SINGLE GFET

| V | I(uA) | R(ohm) |
|---|---|---|
| -1 | -534.8 | 1,869.858 |
| -0.9 | -481.1 | 1,870.713 |
| -0.8 | -427.9 | 1,869.596 |
| -0.7 | -374.4 | 1,869.658 |
| -0.6 | -320.8 | 1,870.324 |
| -0.5 | -267.3 | 1,870.557 |
| -0.4 | -213.3 | 1,875.293 |
| -0.3 | -160.2 | 1,872.659 |
| -0.2 | -106.4 | 1,879.699 |
| -0.1 | -53.1 | 1,883.239 |
| 0.0 | 0.1 | 1,883.000 |
| 0.1 | 53.6 | 1,865.672 |
| 0.2 | 106.9 | 1,870.907 |
| 0.3 | 160.1 | 1,873.829 |
| 0.4 | 213.3 | 1,875.293 |
| 0.5 | 266.5 | 1,876.173 |
| 0.6 | 319.5 | 1,877.934 |
| 0.7 | 372.5 | 1,879.195 |
| 0.8 | 425.6 | 1,879.699 |
| 0.9 | 478.5 | 1,880.878 |
| 1 | 531.5 | 1,881.468 |

| R prom | 1,875.031 |
|---|---|
| R std | 5.173 |

TRIPLE GFET

| V(V) | I(uA) | R(ohm) |
|---|---|---|
| -1 | -926 | 1,079.914 |
| -0.9 | -831 | 1,083.032 |
| -0.8 | -738 | 1,084.011 |
| -0.7 | -645.1 | 1,085.103 |
| -0.6 | -552.9 | 1,085.187 |
| -0.5 | -460.6 | 1,085.541 |
| -0.4 | -368.4 | 1,085.776 |
| -0.3 | -276.2 | 1,086.169 |
| -0.2 | -183.9 | 1,087.548 |
| -0.1 | -92 | 1,086.957 |
| 0.0 | 0.3 | 1,086.000 |
| 0.1 | 92 | 1,086.957 |
| 0.2 | 183.4 | 1,090.513 |
| 0.3 | 275.1 | 1,090.513 |
| 0.4 | 366.8 | 1,090.513 |
| 0.5 | 458.4 | 1,090.750 |
| 0.6 | 550 | 1,090.909 |
| 0.7 | 641.7 | 1,090.852 |
| 0.8 | 733 | 1,091.405 |
| 0.9 | 825 | 1,090.909 |
| 1 | 916. | 1,091.703 |

| R prom | 1,087.631 |
|---|---|
| R std | 3.294 |

MANUFACTURER'S DATA (VACUUM)

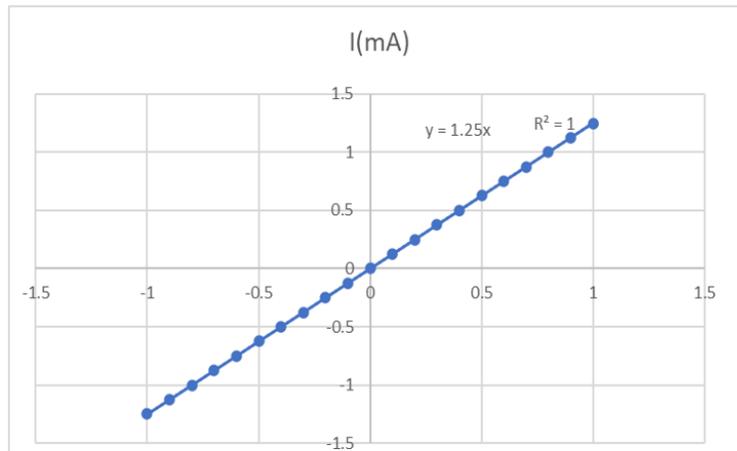

AT ATMOSPHERIC PRESSURE – SINGLE CHANNEL GFETS

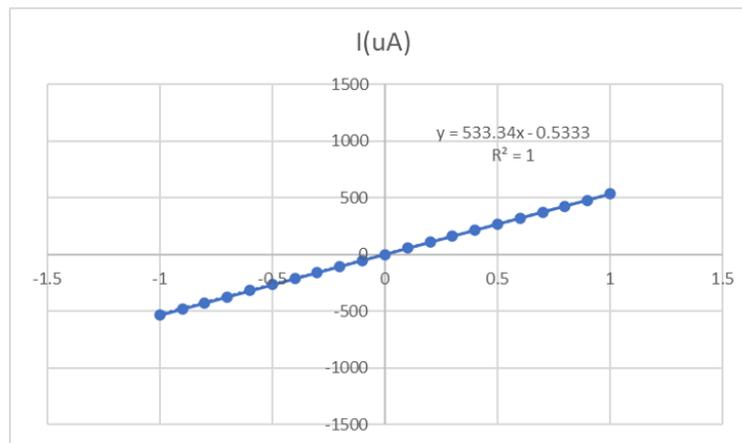

AT ATMOSPHERIC PRESSURE – TRIPLE CHANNEL GFETS

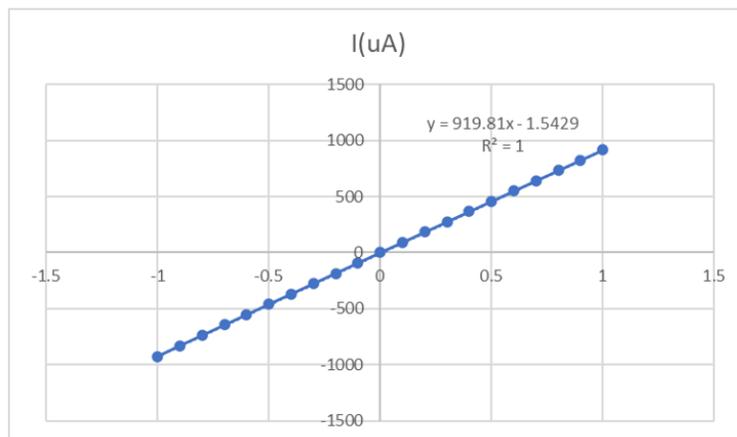

Figure 6. Id-Vd Characteristic curve for channel resistance measurement comparing between single and triple channel devices at atmospheric pressure. Up – data provided by the manufacturer, in the middle – for the single-channel transistor, down – for the triple-channel transistor. (Own elaboration).

Table 2. Measurement of drain current vs. Vds for single channel transistors:
Air dry gate and Gate with PBS solution 0.001X. (Source: authors).

| Chip No. SINGLE GFET | DRY GATE | | Chip No. SINGLE GFET | WET PBS 0.001X | |
|---|---|---|---|---|---|
| V | I(uA) | R(ohm) | V(V) | I(uA) | R(ohm) |
| -1 | -534.8 | 1,869.858 | -1 | -154.0 | 6,493.51 |
| -0.9 | -481.1 | 1,870.713 | -0.9 | -138.0 | 6,521.74 |
| -0.8 | -427.9 | 1,869.596 | -0.8 | -122.0 | 6,557.38 |
| -0.7 | -374.4 | 1,869.658 | -0.7 | -105.8 | 6,616.26 |
| -0.6 | -320.8 | 1,870.324 | -0.6 | -89.9 | 6,674.08 |
| -0.5 | -267.3 | 1,870.557 | -0.5 | -74.3 | 6,729.48 |
| -0.4 | -213.3 | 1,875.293 | -0.4 | -59.0 | 6,779.66 |
| -0.3 | -160.2 | 1,872.659 | -0.3 | -43.8 | 6,849.32 |
| -0.2 | -106.4 | 1,879.699 | -0.2 | -28.9 | 6,920.42 |
| -0.1 | -53.1 | 1,883.239 | -0.1 | -14.4 | 6,944.44 |
| 0 | 0.1 | 1,883.000 | 0 | 0.0 | 0.00 |
| 0.1 | 53.6 | 1,865.672 | 0.1 | 14.8 | 6,756.76 |
| 0.2 | 106.9 | 1,870.907 | 0.2 | 29.2 | 6,849.32 |
| 0.3 | 160.1 | 1,873.829 | 0.3 | 42.9 | 6,993.01 |
| 0.4 | 213.3 | 1,875.293 | 0.4 | 56.8 | 7,042.25 |
| 0.5 | 266.5 | 1,876.173 | 0.5 | 70.5 | 7,092.20 |
| 0.6 | 319.5 | 1,877.934 | 0.6 | 84.2 | 7,125.89 |
| 0.7 | 372.5 | 1,879.195 | 0.7 | 97.6 | 7,172.13 |
| 0.8 | 425.6 | 1,879.699 | 0.8 | 112.0 | 7,142.86 |
| 0.9 | 478.5 | 1,880.878 | 0.9 | 125.0 | 7,200.00 |
| 1 | 531.5 | 1,881.468 | 1 | 139.2 | 7,183.91 |

| R prom | 1,875.031 |
|---|---|
| R std | 5.173 |

| R prom | 6,554.50 |
|---|---|
| R std | 1,519.09 |

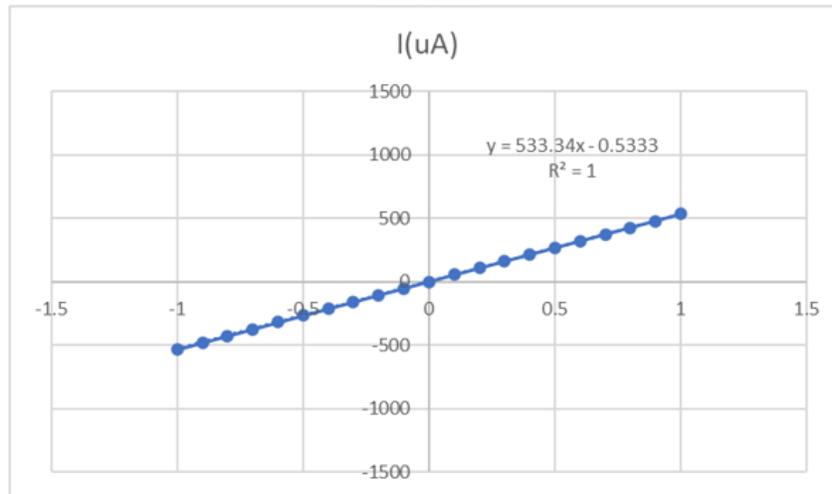

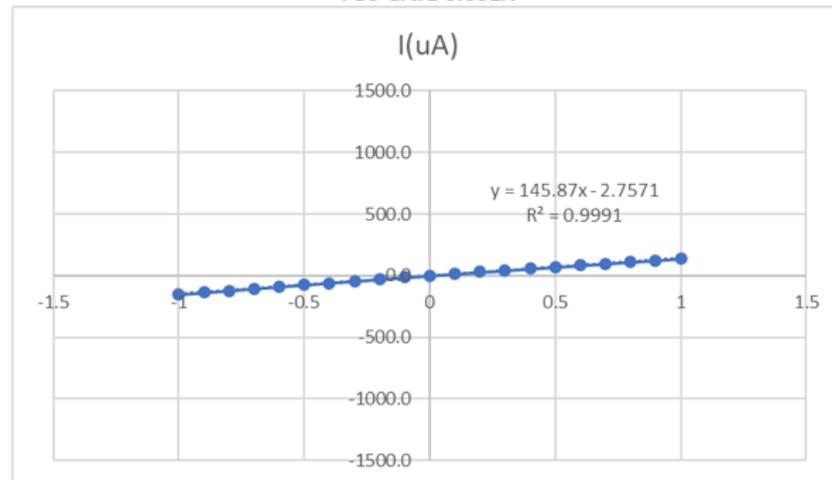

Figure 7. Id-Vd Characteristic curve for channel resistance measurement in single-channel GFET between: Up – with dry gate, down – with wet gate with PBS 0.001X solution, at atmospheric pressure. (Own elaboration)

## 4     CONCLUSIONS AND RECOMMENDATIONS

It was possible to characterize the channel resistance for single and triple channel GFETs using the proposed methodology and workstation. Differences were observed both for the exposure surface (1 vs 3 channels) as well as the presence or absence of a solution in the channel.

Linear models describe the data very well, however, there seems to be a nonlinearity in direct relation to the increase in channel voltage that is more evident in the case of the wet gate. Even in this case, the goodness of fit was very good (R=0.99).

Except for the fact of the equivalent resistance for the case of triple channel that is not 1/3 of that of the single channel, all other behaviors are those expected by the literature. Because it is intended for both alleys of the research to use this device to measure as a biosensor, an important series of recommendations to follow are necessary.

It is recommended that all the GFETs in the arrangement be evaluated, and for each of them, make at least 10 measurements to establish if the manufacturing process is uniform and consequently consider a parameter calibration procedure before making the measurements.

Along with making the statistics for all GFETs, it is recommended to deepen the increasing behavior of resistance with voltage and describe nonlinearity, to contribute in the future with the models of transistors and to be able to consider this deviation when performing analyte detection.

It is important to do the characterization with different concentrations of solution in the gate, as well as bias with a gate-source voltage and locate the Dirac voltage.

It is suggested to complement the characterization with tests in alternating current since the use of liquid in the gate implies important capacitances between the terminals that could vary with the concentration of the analytes and thus complement the DC tests carried out in this work.

Future work in this line should consider studies that incorporate the effects of sterilization techniques in conditions compatible with this type of devices (Gonzalez-Lizardo et al., 2024) not only because of the tests themselves, but also because the same sensor is intended to be used several times for biological analytes and the absence of pathogens is a high priority.

**ACKNOWLEDGEMENTS**

This work was partially funded by the program "PROMOTING POSTBACCALAUREATE OPPORTUNITIES FOR HISPANIC AMERICANS" (PPOHA) PEER TO PEER COLLABORATION FELLOWSHIP – SPRING 2023, promoted by the PUPR Graduate School and VITASENTI LLC.